\documentstyle[12pt]{article}
\begin{document}
\thispagestyle{empty}
\begin{flushright}
{\large \bf JINR preprint E2-92-479\\
Dubna, 1992}
\end{flushright}
\vskip0.8cm
\begin{center}
{\Large \bf On the model of the relativistic particle with curvature and
torsion}\\[0.2cm]
{\large \bf V.\ V.\ Nesterenko}\\
{\it Laboratory of Theoretical  Physics, Joint Institute for
Nuclear Research, SU-141980 Dubna, Russia}\\
\vspace*{0.1cm}
             E-mail address: nestr@theor.jinrc.dubna.su\\[0.5cm]

{\large \bf Abstract }  \\
\vspace*{0.5cm}
\end{center}

Two integrals along the world trajectory
of its curvature and torsion are added to the standard action
for the point-like spinless relativistic particle.
Since here the three-dimensional
space-time is considered at the beginning, the torsion of the world curve is
defined with a
sign in contrast to the previous consideration: {\it V.\ V.\ Nesterenko, J.
Math. Phys. {\bf 32}
3315 (1991)}. Upon obtaining a complete set of  constraints in the phase space
a generalized Hamiltonian description of a new version of the model is
constructed. This enables
one to quantize the model canonically and  to derive exactly the relation
between
the spin and  mass of the states.\\[2cm]

\newpage
\vspace*{2cm }

\section{INTRODUCTION}
Recently the  interest has  been revived  in constructing  and investigating
new models of point-like relativistic particles. To a considerable extent this
is due to the development of the string approach to the unification  of all
the fundamental  interactions. The  straightforward way for obtaining new
particle models is pure geometrical.  The integrals along the particle
world curve of its higher geometrical invariants  should  be added to
the standard action  of the             relativistic
particle [1-6]. The most general model of this kind in
the $D$-dimensional space-time has been investigated in
paper [1] (see also ref.\ [5]). It was the so called  relativistic particle
with curvature and torsion. Inclusion  into the Lagrangian
of higher derivatives with respect to time of dynamical variables entails the
account of  additional degrees of freedom. As a result,
it turns out  well in this approach to describe particles with a nonzero spin
without
introducing additional spin variables.

In paper [1] a generalized Hamiltonian dynamics for a relativistic particle
with curvature and
torsion was  constructed at the very beginning in the $D$-dimensional
space-time $(D\geq 3)$ and only then it was put $D=3$ to simplify quantization.
In
the present paper the same model is formulated at once in the three-dimensional
space-time, the possibility
to define the torsion of the world curve with a sign for $D=3$ being used.
Quantization of this version
of the relativistic particle  with curvature and torsion results in the
spectrum of states which is a linear counterpart of the spectrum obtained in
[1]. The layout
of the paper is as follows. In Sec.\ II the action of a relativistic particle
with curvature and torsion is discussed for arbitrary space-time
dimension $D$ and for $D=3$. In the latter case the torsion of a world curve
is defined with a sign. Here the basic results of paper$^1$ are also presented.
In Sec.\ III a generalized Hamiltinian formalism for a new version of a
relativistic particle
with curvature and torsion is developed ($D=3$ and the torsion is defined with
a sign).
In Sec.\ IV the mass-spin relation is derived and the canonical quantization of
the model
is briefly discussed. In Sec.\ V the obtained results
are compared with the other treatment of this problem.
\section{ACTION OF A RELATIVISTIC PARTICLE
WITH CURVATURE AND TORSION}
\setcounter{equation}0
In paper [1] the following action was considered in the $D$-dimensional
space-time
\begin{equation}
S\,=\,-\,m\int ds\,-\,\alpha\int k(s)\,ds \,-\,\beta \int \kappa (s)\,ds\,{,}
\end{equation}
where $k(s)$ and $\kappa (s)$ are respectively the first and  second
curvatures of the world trajectory of the particle. When $D=3, \quad k(s)$  is
called usually the curvature of the world curve and $\kappa (s)$ is its
torsion.
The model defined by the action (2.1) will be   for simplicity referred to as a
 particle with curvature
and torsion.

As it is known from elementary differential geometry [7], the torsion is
defined at
any point of the curve $x^\mu (s)$ where its curvature
\begin{equation}
k^2\,=\, \frac{d^2x_\mu}{ds^2}\,\frac{d^2x^\mu }{ds^2}
\end{equation}
does not vanish
\footnote{The Lorentz metric with a signature $(+,\,-,\, \ldots \,,\,-)$ is
used.}.
Here $s$ is the natural parameter along the curve, that is,
its length $(dx_\mu / ds)^2=1$. For arbitrary $D > 3$ the second curvature
(torsion) $\kappa (s)$ is defined as follows
\begin{equation}
\kappa \,=\,\frac{\sqrt{\det (d_{\alpha \beta })}}{k^2(s)}\,{,}
\end{equation}
$$
 d_{\alpha \beta}=\stackrel{(\alpha )\,(\beta )}{x^\mu  x_\mu },
\quad   \stackrel{(\alpha)}{x}  \equiv d^\alpha x/ds^\alpha,
\quad\alpha,\beta=1,2,3.
$$
Obviously the function $\kappa (s)$ by  its definition is positive definite.
 In the case of plane curves $(D=2)$ torsion equals zero identically. When
$D=3$, formula (2.3)
has also meaning but in this case there is a possibility to define the torsion
of
a curve with a sign
\begin{equation}
\kappa(s)\,=\, k^{-2}\varepsilon _{\mu \nu \rho}x'^\mu x''^\nu x'''^\rho ,
\end{equation}
where $\varepsilon _{\mu \nu \rho}$ is a completely antisymmetric tensor of the
third rank,
$\varepsilon _{012}=+1$, the prime denotes the differentiation with respect to
$s$.
Thus the torsion of a curve introduced by (2.4) is proportional to a mixed
product of three vectors $x'_\mu ,\; x''_\nu , \;x'''_\rho $. Torsion defined
by (2.4) may differ from (2.3) only by sign minus
at certain points of a curve.

In paper [1] the generalized Hamiltonian description of the
action (2.1) was constructed for arbitrary $D$ and the canonical quantization
was accomplished for $D=3$. The torsion of the world curve was defined by (2.3)
both in the case
of arbitrary $D>3$ and for $D=3$.
The extension of the Dirac theory of  Hamiltonian systems with constraints
to the Lagrangians with higher derivatives developed in paper [2] was
essentially used there. When quantizing this model, natural requirements
for the state vectors $| \psi >$ were  introduced. They should be eigenvectors
of the Casimir operators
$P^2$ and $W$ of the  Poincar\'e group, where $P^\mu $ is the total
energy-momentum vector
and $W$ is a generalization to the arbitrary $D$ of the Pauli-Lubanski vector
squared (spin squared) that is introduced when $D=4$
\begin{equation}
W \,=\,\frac{1}{2}\,M_{\mu \nu }M^{\mu \nu}P^2\,-\, (M_{\mu \nu}P^\nu )^2.
\end{equation}
Here $M_{\mu \nu}$ are the generators of the Lorentz
group. Thus, the state vectors obey the equations
\begin{eqnarray}
P^2\,|\psi >&=&M^2\,|\psi >,\\
W\,|\psi > &=&M^2j(j+1)\,|\psi >.
\end{eqnarray}
It is essential that the spin of the state $j$ can take arbitrary non-negative
values (integer,
half-odd-integer and fractional, $D=3$).

For $M^2>0$ the mass-spin relation (the Regge trajectory)
\begin{equation}
\left ( \frac{M}{m}\right )^2j^2\,=\,\left ( \alpha \sqrt{1-\left (\frac{M}{m}
\right )^2}\,+\,|\beta |\right )^2
\end{equation}
was derived.

  The aim of the present paper is the investigation of the model (2.1) in the
three-dimensional space-time by making use of the torsion definition
given by (2.4). It should be noted at once that in this case space symmetry of
the theory is
violated due to the pseudoscalar nature of  formula (2.4). But it is this model
that arises under
investigation of  charged scalar particles placed in an external abelian
Chern-Simons field [8, 9].
\section{GENERALIZED HAMILTONIAN FORMALISM}
\setcounter{equation}0
Upon introducing the arbitrary parametrization of the world
curve, $x_\mu(\tau ), \quad \mu =0,\,1,\,2$, the action (2.1) with allowance
for (2.4) reads as follows
$$
S\,=\,-m\int d\tau \sqrt{\dot x^2} -  \alpha \int d\tau \frac{\sqrt{(\dot
x\ddot x)^2 - \dot x^2\,\ddot x^2}}{\dot x^2} -
$$
\begin{equation}
-\beta \int d\tau \sqrt{\dot x^2}\,\frac{\varepsilon_{\mu \nu \rho}\dot x^\mu
\, \ddot x^\nu \stackrel{\ldots}{x}^\rho}
{{(\dot x\ddot x)^2\,-\,\dot x^2\,\ddot x^2}},
\end{equation}
$$
\dot x \equiv dx(\tau )/d\tau,\quad D\,=\,3.
$$
The Hamiltonian description of the model (3.1) can be
constructed following the papers [1, 2]. Canonical variables are introduced in
a standard fashion
$$
q_1\,=\,x, \quad q_2\,=\,\dot x, \quad q_3\,=\,\ddot x
$$
\begin{equation}
p_1\,=\,-\,\frac{\partial L}{\partial \dot x}\,-\, \dot p_2,\quad
p_2\,=\,-\,\frac{\partial L}{\partial \ddot x}
\,-\,\dot p_3,\quad p_3\,=\,-\,\frac{\partial L}{\partial
\stackrel{\ldots}{x}}\,{,}
\end{equation}
where $L$ is the Lagrangian function in (3.1). The equations of motion are
reduced to the
obvious conclusion $p_1\,=\,$constant (the conservation law of the
energy-momentum). Therefore the dynamics
of the model under consideration is determined by the constraints.

By making use of the definition
of the canonical momentum $p_3$ and  the exact form of $L$ we derive three
primary
constraints
\begin{equation}
\stackrel{(1)}{\varphi }_\mu \,=\,p_{3\mu }\,+\,\beta
\,\frac{\sqrt{q_2^2}}{g}\,
\varepsilon_{\mu \nu \lambda }\,q_2^\nu \,q_3^\lambda \,=\,0\,{,}
\end{equation}
$$
\mu ,\,\nu ,\, \lambda \,=\,0, \,1,\,2\,{,}
$$
where $g=(q_2\,q_3)^2\,-\,q_2^2\,q_3^2$. After squaring this equation and
projecting it
onto $q_2$ and $q_3$ one could get the  primary constraints identical with
those  used
in paper [1]. However it is important in the following that the primary
constraints are kept in the original form (3.3).

 The canonical Hamiltonian has the same form as in paper [1]
\begin{equation}
H\ =\ -p_1 \,\dot x\,-\,p_2\,\ddot x\,-\,p_3\stackrel {\ldots
}{x}\,-\,L\,=\,-p_1\,q_2\,-\,p_2\,q_3\,+\,
m\,\sqrt{q_2^2}\,+\,\alpha \,\frac {\sqrt {g}} { q^2_2} \,{.}
\end{equation}

A complete set of  constraints can be determined by the Dirac method [2]. The
requirement of  stationarity of the primary constraints
results in three secondary constraints
$$
\stackrel{(2)}{\varphi }_\mu \,=\,p_{2\mu }\,-\,\frac{\alpha }{q_2^2\sqrt{g}}\,
[\,(q_2\,q_3)\,q_{2\mu }\,-\,q_2^2\,q_{3\mu }\,]\,-\,
$$
\begin{equation}
-\beta \,\varepsilon_{\mu \nu \lambda }\,q_2^\nu \,q_3^\lambda
\,\frac{(q_2\,q_3)}{g\,\sqrt{q_2^2}}\,=\,
0\,{,}\quad \mu\,=\,0,\,1,\,2\,{.}
\end{equation}
In its turn the stationarity condition for (3.5) entails three new ternary
constraints. They are
rather complicated in form, therefore, it is worthwhile to write them at once
in the proper time gage
\begin{equation}
q_2^2\,=\,\mbox{const}, \quad q_2\,q_3\,=\,0\,{;}
\end{equation}
then one gets
\begin{equation}
\stackrel{(3)}{\varphi }_\mu \,=\,p_{1\mu }\,-\,
m\,\frac{q_{2\mu }}{\sqrt{q_3^2}}\,-\,\beta \,\frac{\varepsilon_{\mu \nu
\lambda }\,q_2^\nu\,q_3^\lambda }
{q_2^2\,\sqrt{q_2^2}}\,=\,0, \quad \mu\,=\,0,\,1,\,2\,{.}
\end{equation}
There are no other constraints in this model. The constraints (3.5) and(3.7)
with
allowance for (3.3) and (3.6) coincide with formulae (3.8) and (3.10) of  paper
[1].
\section{SPIN-MASS RELATION}
\setcounter{equation}0
The canonical momentum $p_{1\mu }$ represents the total energy-momentum vector.
Squaring Eq.\ (3.7) one obtains
\begin{equation}
p_1^2\,\equiv \,M^2\,=\,m^2\,+\,\beta ^2\,\frac{q_3^2}{(q_2^2)^2}\,{,}
\end{equation}
where $M^2$ is the mass of a particle in the model under consideration.
Since $q_3^\mu $ may be time-like, space-like or an isotropic vector it follows
that $M^2$ is not positive definite. By making use of the definition of the
curvature (2.2) Eq.\ (4.1) can be rewritten as
\begin{equation}
p_1^2\,\equiv \,M^2\,=\,m^2\,-\,\beta ^2\,k^2(s).
\end{equation}
Thus the mass  of the state is  expressed in terms of the curvature of
the world curve in the same way as in the former version of this model [3].
{}From (4.2)
it follows in particular that curvature $k(s)$ is a constant (it is an integral
of motion).

If the states are classified under eigenvalues of  the quadratic Casimir
operators
(2.6) and (2.7) as it was done in paper [1], then one arrives at the same mass
spectrum (2.8).
A somewhat different situation will be in the case when the spin operator $S$
given by
\begin{equation}
S\,=\,\frac{1}{2\sqrt{|p_1^2|}}\,\varepsilon_{\mu \nu
\lambda}\,p_1^{\mu}\,M^{\nu \lambda}.
\end{equation}
is used instead of $W$. The spin $S$ is a pseudoscalar quantity and it may take
both positive and negative values.

To derive the mass-spin relation, one has to calculate the spin $S$ on the
submanifold of the phase space defined by the constraint equations (3.3),
(3.5), (3.7),
 and the gauge conditions (3.6). Substituting the Lorentz generators $M_{\mu
\nu}$
\begin{equation}
M_{\mu \nu}\,=\,\sum_{a=1}^{3}(\,q_{a\mu}\,p_{a\nu }\,-\,q_{a\nu }\,p_{a\mu}\,)
\end{equation}
into (4.2) one obtains
\begin{equation}
S\,=\,\frac{1}{\sqrt{| p_1^2|}}\,\varepsilon_{\mu \nu \lambda}\,p_1^\mu
\,(\,q_2^\nu \,p_2^\lambda\,+\,q_3^\nu \,p_3^\lambda \,)\,{.}
\end{equation}
Now the constraints and  the gauge conditions should be taken into account.
Here the manifest form of the
primary constraints is very important. Only in the case when they are taken in
the original form
(3.3) instead of a squared form (see Eqs.\ (2.8) -- (2.10) in paper [1]) the
spin
value on the submanifold introduced above can  be calculated exactly. Equation
(4.5) acquires now the form
$$
S\,=\,\frac{-\,\varepsilon_{\mu \nu
\lambda}}{\sqrt{|p^2_1|}\,q_2^2\,\sqrt{-q_3^2}}\,
\left (\,m\,q_2^\mu \,+\,\frac{\beta }{q_2^2}\,\varepsilon ^{\mu \sigma \rho}
\,q_{2\sigma}\,q_{3\rho }\,\right )
$$
\begin{equation}
\left (\,\alpha \,q_2^\nu \,q_3^
\lambda\,+\,\frac{\beta}{\sqrt{-q_3^2}}\,q_3^\nu \,\varepsilon^{\lambda \rho
\sigma}\,q_{2 \rho}\,q_{3\sigma }\,\right )\,{.}
\end{equation}
After simplification it reads
\begin{equation}
S\,=\,\frac{\beta}{\sqrt{|p_1^2|}}\,\left (\,m\,
+\,\alpha\,\frac{\sqrt{-q_3^2}}{q_2^2}\,\right )\,{.}
\end{equation}
 From (4.1) it follows that
\begin{equation}
|\beta|\,\frac{\sqrt{-q_3^2}}{q_2^2}\,=\,\sqrt{m^2\,-\,p_1^2}\,{.}
\end{equation}
Substitution of Eq.\ (4.8) into (4.7) gives
\begin{equation}
S\,=\,\varepsilon_\beta\,\left (\,\alpha\,\sqrt{\frac{1}{\mu^2}\,-\,\varepsilon
}\,
+\,\frac{|\beta|}{\mu}\,\right )\,{,}
\end{equation}
where $\varepsilon_\beta\,=\,\mbox{sign} \,\beta,\quad
\varepsilon\,=\,\mbox{sign} \,p_1^2,\quad \mu \,
=\,\sqrt{|p_1^2|}/m\,\leq\,1$. At certain values of the parameters $\alpha $
and $\beta$ the
right hand side in Eq.\ (4.9) considered as a function of $\mu $ has an
extremum. For example,
if $\beta ^2\,>\,\alpha ^2$ and $\varepsilon \,=\,1$ this takes place at the
point $\mu \,=\,
\sqrt{\beta ^2\,-\,\alpha ^2}/|\beta|$. Near by this point two values of the
state mass $\mu $
correspond to the same spin value $S$. Thus, there is a "degeneracy" of the
mass spectrum with respect to spin. The analogous
property is encountered in the theory of  infinite component
relativistic  wave equations [10].

Squaring the left- and right-hand sides of Eq.\ (4.9) one gets, as it could be
expected,
the mass-spin relation (2.8). All, that concerns the double-valued property of
the spectrum (4.9), is
valid obviously for Eq.\ (2.8) too.
In order to reveal the double-valued behavior of Eq.\ (2.8) it was resolved in
paper [1]
with respect to $\mu $ but unfortunately not quite correctly. The right formula
has rather
a cumbersome form
$$
\mu \,=\,\frac{\left|\,j\,|\beta
|\,\pm\,\sqrt{\alpha^2\,(j^2\,+\,\alpha^2\,-\,\beta^2)}\,\right|}
{\alpha ^2\,+\,j^2}\,{,}\quad j\, \geq \,0\,{.}
$$
For each of two values of $\mu $ the corresponding conditions
$$
\mbox{sign}\,\alpha\,=\,\mbox{sign}\,\left (-|\alpha \,\beta |\,\pm
\,j\,\sqrt{j^2\,+\,\alpha ^2\,-\,\beta ^2}\,\right),
\quad \mu\,\leq\,1
$$
should be satisfied.

The quantization of the model retains the mass-spin relation (4.9) for the
eigenvalues of the spin operator $S$
and the mass squared operator $p^2_1$. In the rest frame ($p_1^0\,=\,M,\quad
{\bf p}_1\,=\,0$) the
spin operator $S$ is
\begin{equation}
S\,=\,M^{12}\,{.}
\end{equation}
Without loss of  generality it can be realized as follows
\begin{equation}
S\,=\,-i\,\frac{\partial}{\partial \varphi}\,+\,c\,{,}
\end{equation}
where $\varphi$ is an  angular variable and $c$ is a constant to be determined
below. As a wave
function we take $2\pi $-periodic functions
\begin{equation}
\psi (\varpi)\,=\,\sum_{l\ni Z}^{}\,e^{il\varphi}a_l\,{.}
\end{equation}
The requirement that $\psi (\varphi )$ should be an eigenfunction of $S$ gives
\begin{equation}
l\,+\,c\,=\,S\,{,}
\end{equation}
where $S$ is determined in Eq.\ (4.9). Hence $l$ is an integer part of the spin
$S$ and
$c$ is its fractional part, that is, the spectrum of the spin operator is
continuous.
\section{CONCLUSIONS}
As  it should be expected, the allowance for the torsion sign in
 the action (3.1) results in the mass-spin relation (4.9) that can
be treated as the square root of the spectrum (2.8). In  paper [5] the action
(3.1)
has been investigated by eliminating
higher derivatives with the aid of the
corresponding Lagrange multipliers. The spectrum derived there is not identical
with  the spectrum (4.9) though they are alike. Instead of the degeneracy with
respect to spin, the mass-spin relation obtained in [5] gives the  degeneracy
with respect to  mass of a state,
that is, at certain values of the  model parameters
 the states with different values of the spin may have the same mass.\\[1cm]
{\bf ACKNOWLEDGEMENT} \\[0.3cm]

The author is grateful to M.\ S.\ Plyushchay discussions with whom have
assisted in elucidating
the distinctions between the models (2.1) and (3.1).
\newpage
\begin{flushleft}
{\bf REFERENCES}
\end{flushleft}
\begin{enumerate}
\item V.\ V.\ Nesterenko, J.\ Math.\ Phys.\ {\bf 32}, 3315 (1991).
\item V.\ V.\ Nesterenko, J. Phys.A:\ Math.\ Gen. {\bf 22}, 1673 (1989).
\item V.\ V.\ Nesterenko, Class.\ Quantum Grav.\ {\bf 9}, 1101 (1992).
\item V.\ V.\ Nesterenko, Int.\ J.\ Mod.\ Phys.\ {\bf A6}, 3989 (1991).
\item Yu.\ A.\ Kuznetsov, M.\ S.\ Plyushchay
(preprint IHEP 91 -- 162), Protvino,1991; Nucl.\ Phys.~B.
\item Yu.\ A.\ Kuznetsov, M.\ S.\ Plyushchay (preprints IHEP 92 -- 96,
92 -- 108), Protvino, 1992.
\item M.\ DoCarmo, {\it Differential Geometry of Curves and Surfaces}
(Printice-Hall,
Englewood Cliffs, NJ, 1976).
\item A.\ M.\ Polyakov, Mod.\ Phys.\ Lett.\ {\bf 3A}, 325 (1988).
\item A.\ L.\ Kholodenko, Ann.\ Phys.\ {\bf 202}, 186 (1990).
\item A.\ O.\ Barut and R.\ Raczka, {\it Theory of Group Representations
and Applications} (Polish Scientific Publishers, Warszawa, 1977).
\end{enumerate}
\vfill
\begin{center}
Received by Publishing Department\\
on November 20, 1992
\end{center}
\end{document}